\documentstyle[prb,aps,psfig,multicol]{revtex}
\begin{document}
\draft
\tighten

\title{The effect of strain on the adsorption of CO on Pd(100)}
\author{M. W. Wu and H. Metiu$^*$}
\address{Department of Chemistry and Biochemistry,
 University of California, Santa Barbara, CA 93106}

\date{\today}
\maketitle
\begin{abstract}
We study how the binding energy, the vibrational frequencies and the adsorption 
isotherm of CO on
Pd(100) are  modified when the solid is subject to uniform strain.  The
parameters controlling the thermodynamics of adsorption (the
adsorption energy,
the vibrational frequencies of the adsorbed molecules and the interaction
energy between the molecules) have been calculated by
using density functional
theory. These parameters are then used in a grand canonical Monte Carlo
simulation that determines the CO coverage when the surface is in
thermodynamic equilibrium with a CO gas, at a specified pressure and temperature.    We
find that this is substantially affected by the strain. Our results, along with those obtained 
by others,  suggest
that the development of  ``elastochemistry", a study of the change in the
chemical properties of a surface when subjected to strain, will lead to
interesting and measurable results.  It also suggests that differences in
chemical activity between clusters on a support
and clusters in gas phase may
be partly due to the strain induced when a cluster is placed on the  support.
\end{abstract} 

\begin{multicols}{2}
\narrowtext

\section{Introduction}
A number of recent experiments and calculations have shown that
chemisorption causes  surface strain and that straining a surface
modifies the properties of adsorbed molecules. Webb, Lagally and
their students \cite{webb,lagally}
demonstrated that the structure of a silicon surface can
be changed by bending it. They worked with a (100) face  of a
thin  silicon slab. This surface has steps separated by two kinds
of terraces: on  one,  the dimer rows are parallel to the steps; on
the other, they are perpendicular to the steps. These terraces give
different  low electron diffraction (LEED) patterns. When the slab is
bent, the relative LEED intensities  change, because the size of
these terraces is modified by strain. Ibach and his
 coworkers\cite{ibach1,ibach2,ibach3,ibach4} have demonstrated a
``reciprocal'' phenomenon: chemisorption  on the surface of a very
thin  slab will cause it to bend. If adsorption causes strain, then
it should be possible to affect adsorption by straining the surface.
Mentzel's group\cite{menzel} implanted noble gas ions under a
metal surface, causing local strain. They then showed that the  strained  
surface has  different adsorption properties.  Other
groups\cite{goodman,madey,behm,kern,chork,besenbacher} have strained a surface 
by growing a very thin
metal film on a substrate of a different metal. If the two lattices are
mismatched,  but the growth is in registry,  the atoms in the film
are either stretched or compressed. This affects the chemisorption
properties of the film. In this kind of experiment the
chemisorption properties are modified by the strain and also by electronic
effects caused by binding to a substrate made of a different metal. The
two effects cannot be separated experimentally.

Density functional calculations have been used to explore the effect
of surface strain on the properties of the adsorbates. Ratsch, Seisonen and 
Scheffler\cite{scheffler} have shown
that the activation energy for diffusion of a Ag atom on a Ag surface is
affected by strain. Mattsson and Metiu\cite{metiu1} have used this effect
to show that periodic strain on a surface\cite{petroff} can order
nanostructures nucleated on it and increase their size uniformity.
Finally, Mavrikakis, Hammer and N\o rskov\cite{norskov} calculated the change
in the binding energy of O and CO  and in  the dissociation energy of CO
when the  Ru(0001) surface on which they are adsorbed is under strain.

Ibach\cite{ibachreview} has published a thorough review of the
effects of strain in epitaxy and surface reconstruction.
Norskov\cite{norskovreview} discusses the effect of strain in
an excellent review of density functional studies of chemisorption systems
relevant to catalysis.

It is not surprising  that straining a surface  affects the
chemistry taking place on it. Strain changes the distance between the
surface atoms and this must change the properties of the adsorbates. The
only question is whether this change can be, at least in some cases,
sufficiently large to matter. If this is so, one can envision the
development of ``elastochemistry'' as a new subfield of physical
chemistry. 

In this paper we study how the binding energy, the vibrational frequencies
and the adsorption isotherm of CO on Pd(100)  change with surface strain
and coverage. We use  generalized gradient, density  functional  theory
for energy calculations  and Monte-Carlo simulations and an analytical
model for calculating the adsorption isotherm. The change in these
quantities, caused by straining the surface, is sufficiently large to be
measurable. In the case of the adsorption isotherm the coverage
at a given gas pressure changes by almost two orders of magnitude. This
happens because relatively small changes in adsorption energy have large
effects on the isotherm. 

To model the  adsorption isotherm we need to know how the properties
of the adsorbed molecules change with coverage. Ideally, we should calculate
how these properties change when we change the clustering of CO on the
surface. One would need then the energy of all possible clusters. This is
beyond the current capability of the density functional method, which uses a
periodic system with a unit cell of limited size. For this reason
we have adopted the following strategy.  We calculate the properties of
adsorbed CO  for 1/8, 1/2 and 1 monolayer. The
one-monolayer calculation is used to determine the adsorption energy of
a molecule that does not interact with its neighbors. The half-monolayer
results provide the interaction energy between the next-nearest-neighbors. The
full monolayer one is used to extract the interaction energy between
nearest-neighbors. A Hamiltonian based on these energies is
then used in a grand-canonical Monte Carlo simulation to determine the
adsorption isotherm. We have also tested  the quasi-chemical 
approximation\cite{hill} and found it to be inaccurate.
\begin{figure}[htb]
\psfig{figure=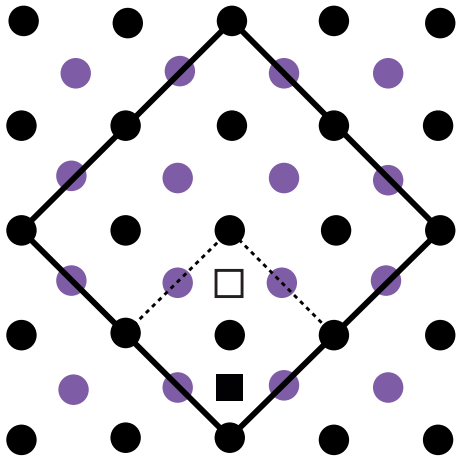,width=6.cm,height=6.cm,angle=0}
\bigskip
\caption{The square drawn with  thick lines shows the supercell used
in low CO coverage (1/8\ ML) calculations. The dashed lines show the
supercell used for calculations in which CO coverage is  1 or  1/2\ ML.
Dark dots: Pd atoms in the top layer; Gray dots: Pd atoms in the
subsurface layer.  Black and open squares: the bridge sites occupied by
CO. If the coverage is 1/8\ or 1/2\ monolayer only the dark square is
occupied. Both squares are occupied if the coverage is 1 monolayer.}
\end{figure}

\section{Methodology}
The density functional calculations reported here were performed with the Vienna  program VASP.\cite{kress1,kress2} This uses periodic
boundary conditions, a plane-wave basis set and fully nonlocal
Vanderbilt-type \cite{van} ultra soft psudopotentials. The generalized gradient 
exchange-correlation energy is that of Wang and Perdew\cite{wang}.

The metal is represented by  a four-layer slab bounded by an empty
region whose thickness equals that of  six Pd layers.  When the coverage 
 $\rho$  is
high ($\rho=1$ and 1/2), we use the small supercell shown by the
 dashed lines in Fig.\ 1; the thick lines show the  supercell for the lowest coverage  ($\rho=1/8$).

The  CO molecules are placed on the top layer of the slab.  The positions
of the atoms in the bottom layer was fixed to coincide with those of
the bulk Pd,  calculated with the same program.
The distance between the bottom
layer and the layer next to it is also fixed to the bulk value.
The calculated  equilibrium lattice constant of bulk Pd is
$a=3.961$\ \AA. The atoms in the other layers are  allowed to relax and
take the values that give a minimum total energy.

Brillouin-zone integrations have been performed on a grid of
$9\times 9 \times 1$ ${\bf k}$ points for the high-coverage case and
on a grid of $4\times 4\times 1$ ${\bf k}$ points for the low-coverage case. The Methfessel-Paxton smearing is  $\sigma=0.3$\ eV. 
For high coverage we have tested convergence by performing calculations 
with thicker slab and a denser ${\bf k}$ mesh.  For low coverage, a
$4\times 4\times 1$ ${\bf k}$-point mesh is the densest we can
afford on a workstation with 1GB memory. The cut-off energy is  495\ eV,
which is a high value.

\section{The properties of adsorbed CO, as a
function of coverage and strain} 

\subsection{Bulk Pd}

To determine the lattice constant of bulk Pd we used the supercells described in Fig. 1 and a slab having ten Pd layers. It is not necessary to use such a thick slab, but we had the results from calculations for another project.
If the sample is unstrained,  we allow the atoms to relax in all directions
and obtain a lattice constant  of  3.961\ \AA. This corresponds to a
distance of  1.9805\ \AA\  between the (100) layers.  We have also
calculated the bulk structure of two samples, subject to  an uniform strain
of $\pm2\%$ in a plane  (the xy plane)  perpendicular to the (100)
direction. In these calculations the $x$ and $y$ coordinates of the atoms
are fixed, but the space between the layers is  allowed to relax. In
the sample with $2\ \%$ strain the distance (in the z direction)
between the layers is  1.9250\ \AA; if the strain is  $-2$\%, this distance
is 2.0457\ \AA. Increasing the distance between atoms in the $xy$
plane decreases the distance between the layers in the $z$ direction.

\subsection{Strain and coverage dependence of adsorption energy}
All calculations involving adsorbed molecules are performed on a slab
having $ 4$ Pd  layers, surrounded by a vacuum whose dimension in
the z-direction is equal to that occupied by $6$ Pd layers.  The distance
between the lowest two layers is
fixed at the bulk value mentioned above (this distance depends on the
state of strain of the surface). The top two layers are allowed to relax.
We calculate the energy $E_s$ of the slabs constrained in this way for
0  and  $\pm 2$\% strain.

The properties of the chemisorbed CO molecules are calculated by placing 
1 ML, 1/2 ML and 1/8 ML CO on top of the Pd slab.  Since these calculations are
extremely time consuming, we have investigated only the bridge binding
sites, which have the lowest binding energy when the surface is not under strain.\cite{eichler}
It is
possible that other sites are occupied with a finite probability, when the system is in 
thermal equilibrium.  Also, the CO molecules may prefer a different binding site, when 
the surface is under strain. These
possibilities have not been considered in our calculations.
We also ignore the fact that at high coverage the CO molecules might change
their binding site.

 The energy of the CO covered slab is denoted by $E_{tot}$,
and depends on CO coverage and on surface strain. The adsorption energy 
$\overline{\varepsilon}_{ad}$ is then given by: 
\[\overline{\varepsilon}_{a}=(E_{tot}-E_s-nE_{CO})/n\ .\]
$E_{CO}=-14.73$ meV is the total energy of a
free CO molecule, $n$ is  the number of CO molecules in a supercell and
$E_s$ is the energy of the Pd slab, without CO on it.  For the supercells
used in our calculations, $n=2$ for 1 ML of CO and $n=1$ for 1/2 ML or
1/8 ML (see  Fig.\ 1).
The adsorption energies calculated with this equation,
for different CO coverages and strains,  are given in Table I. For
1/2\ ML and no strain, the adsorption energy of  $-1.918$\ eV is  consistent
with that  reported in Ref.\ \onlinecite{eichler}, which is $-1.92$ eV.

The adsorption energy $\bar{\varepsilon}_a$ defined above  includes the
lateral interactions 
between the molecules. Such a definition is often used when the dependence
of desorption rate on coverage is modelled.  As we shall see later,
this definition does not work well; we prefer the one given when we define  
Model\ II (see  Section D).

\subsection{The dependence of the vibrational frequencies of CO on coverage
and strain}

The dependence of the vibrational frequencies of chemisorbed CO on
strain is of interest for two reasons. They are measurable and
therefore the predictions made here can be tested by experiment. Furthermore,
the change of these frequencies with strain will affect the chemical
potential of the chemisorbed molecules and therefore will influence the
adsorption isotherm. A thorough study of these effects would have to
perform a complete phonon analysis of the system.  Such calculations
are possible, in principle, but  they require too much computer power, especially
for the case of low coverage, to be attempted here. Since the adsorption
isotherm is dominated by the chemisorption energy and by the lateral
interactions, small errors in evaluating the vibrational energies are
not likely to affect our conclusions.
\begin{table}[htb] 
\caption{$\overline{\varepsilon}_{a}$ for
different coverage and strain (unit: eV)}

\begin{tabular}{cccc}
STRAIN & 1/8 ML & 1/2 ML& 1ML  \\
\hline
 $-2$\ \% & $-2.203$ & $-1.8723$&$-1.20399$\\
\hline
0\ \%& $-2.228$ & $-1.918$ & $-1.3076$\\
\hline
 2\ \% & $-2.255$ & $-1.944$ & $-1.3826$\\
\end{tabular}
\end{table}

Because of these considerations, we calculate the vibrational frequencies by
giving small displacements to the O and C atoms, away from their
equilibrium positions, while keeping the metal atoms fixed. The
displacements along the $x$ and
$y$ directions are  $\pm 0.05$\AA\  and $\pm 0.1$\AA , and those along
the $z$ direction are
$\pm 0.01$\AA\  and $\pm 0.02$\AA.  We calculate the forces exerted on the C and O atoms, 
for all combinations of
shifted coordinates, and fit it to a linear form in the
displacements. This allows us to calculate the force constants  and then the
vibrational frequencies. The results are given  in Table II, for different
coverages and strains.   In Fig.\ $4$ we show the amplitudes of four vibrations in the
 $xy$ plane, as seen by looking down towards the
surface. The highest frequency, $\omega_6$, is the CO stretch. In
the absence of strain, at a coverage of 1/2 ML, our result (1878 cm$^{-1}$)
is close to that calculated by Eichler and Hafner\cite{eichler}
(1887 cm$^{-1}$). They also report a frequency of 417 cm$^{-1}$ for 
the carbon-metal stretch; our calculations give  341 cm$^{-1}$.
This discrepancy  is  not
sufficiently large to affect the chemisorption isotherm. We suspect that this is due to
 the fact that we use  more {\bf k} points in our calculation.

The frequency $\omega_6$ is  easiest to measure. Our calculations predict
that either a compressive or a tensile strain, which changes the lattice
constant by two percent, will increase $\omega_6$ by an amount measurable
by infrared spectroscopy. The largest change, of 29 cm$^{-1}$, is
at the lowest coverage (1/8 ML). 
\end{multicols}
\widetext
\vskip -0.6cm
\begin{table}[htb]
\caption{Frequencies of CO for different coverage and strain (unit: 1/cm)}

\begin{tabular}{cccccccc}
Coverage &STRAIN & $\omega_{1}$ &$\omega_{2}$ & $\omega_{3}$ &
$\omega_{4}$& $\omega_{5}$ & $\omega_{6}$  \\
\hline
1/8 ML& $-2$\ \% & 397.42 & 334.23 & 169.51 & 49.94 & 338.13 & 1849.71\\
\hline
1/8 ML& 0\ \% & 416.01 & 346.05 & 180.26 & 55.86 & 334.00 & 1820.05\\
\hline
1/8 ML& 2\ \% & 422.36 & 327.99 & 182.70 & 72.12 & 319.95 & 1849.80\\
\hline
1/2 ML& $-2$\ \% & 407.33 & 339.38 & 178.37 & 64.27 & 395.00 & 1890.61\\
\hline
1/2 ML& 0\ \% & 406.60 & 330.06 & 170.89 & 14.45 & 340.63 & 1878.08\\
\hline
1/8 ML& 2\ \% & 414.24 & 323.98 & 182.64 & 71.50 & 343.95 & 1883.84\\
\hline
1 ML& $-2$\ \% & 417.83 & 379.86 & 180.94 & 56.17 & 367.33 & 1995.30\\
\hline
1 ML& 0\ \% & 416.85 & 374.93 & 193.24 & 77.14 & 362.00 & 1982.70\\
\hline
1 ML& 2\ \% & 393.38 & 356.56 & 187.32 & 77.00 & 341.90 & 1986.67\\
\end{tabular}
\end{table}
\begin{multicols}{2}
\narrowtext
\begin{figure}[htb]
\psfig{figure=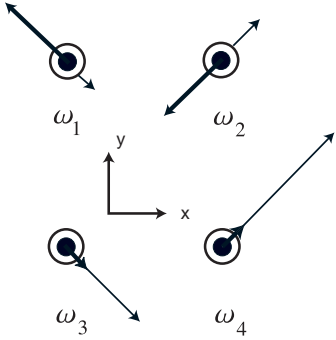,width=7.cm,angle=0}
\vskip -3cm
\caption{Illustration of oscillation modes of CO in $x$-$y$ directions.
Black dot: C; Circle: O.}
\end{figure}

\subsection{The interaction between the adsorbed CO molecules}
To calculate the adsorption isotherm we need to know the
energy of the interaction between molecules. A very precise treatment 
of this problem would have to determine the energy of
dimers, trimers,\ldots. This would require an enormous
amount of computer power and would be an overkill at
this stage in  our research. For this reason we have
studied two approximate models.

In Model I we assume that the effect of the interaction
between the adsorbates is to make the adsorption energy 
depend on coverage. This is obviously a mean field
model; in reality the desorption energy is different, for
different local configurations. For each strain, the adsorption
energy at the three coverages used here is given in Table I.
For other coverages we interpolate between these values. 

In Model II we partition the total energy into an adsorption energy and
the energy of interaction between the adsorbed molecules.   First we
assume that at a coverage of 1/8 ML there
are no interactions between the CO molecules. Therefore,
\[\varepsilon_a=\frac{E_t(1/8)-E_s-n(1/8) E_{CO}}{n(1/8)}\]
gives the adsorption energy
of a CO molecule. Here $E_{t}(1/8)$ is the total
energy of the slab with 1/8 ML CO on
it, $E_s$ is  the total energy of the slab without CO, $E_{CO}$ is the energy of 
the gaseous CO  and 
$n(1/8)$ is the number of CO molecules in the
supercell when the CO coverage is 1/8 ML.

Next, we assume that
the  adsorption energy of one molecule  is independent of
coverage. This means that we attribute the change in adsorption energy
at higher coverages to  the interaction between molecules. When the
coverage is
1/2\ ML we assume that the total energy is the slab energy, plus
the adsorption energy of the molecules in the supercell, plus the interaction
between all the next-nearest-neighbors (nnn) in the supercell. Thus, we
can calculate the interaction $\varepsilon_{nnn}$ between the nnn-pairs
from the equation \[\varepsilon_{nnn}
=\frac{E_t(1/2)-E_s(1/2)-n(1/2) E_{CO}-n_a(1/2)\varepsilon_{a}}{n_{nnn}(1/2)}\ .\]

Here $n_a$ is the number of adsorbed CO molecules in the unit cell.  The total energy of a  CO  monolayer can then be used
to calculate the interaction energy between the nearest neighbors
 (nn). We assume that the total energy of the CO-covered surface is 
the energy of the slab, plus the adsorption energy of the CO molecules, 
plus the interaction energy of all the nn and  nnn pairs in the supercell.
As a consequence, the interaction energy $\varepsilon_{nn}$
between the nearest neighbors is given by
\[\varepsilon_{nn}=\frac{E_t -E_s-n_{nn}(1) E_{CO}-n_{nnn}(1)\varepsilon_{nnn}-
n_a(1)\varepsilon_a}{n_{nn}(1)}\]

The values of $\varepsilon_{nn}$ and $\varepsilon_{nnn}$
calculated from these equations, for the strained and the unstrained
surfaces, are given in Table\ III\@. 
\begin{table}[htb]

\caption{$\varepsilon_{nn}$ and $\varepsilon_{nnn}$ dependence on strain}

\begin{tabular}{ccc}
Strain&$\varepsilon_{nn}$&$\varepsilon_{nnn}$\\
\hline
$-2$\ \%&$0.334$&$0.165$\\
\hline
$0$\ \%&$0.305$&$0.155$\\
\hline
$2$\ \%&$0.2805$&$0.1555$\\
\end{tabular}
\end{table}

We have now all the elements needed for
 writing an effective 
Hamiltonian to be used in the statistical mechanical theory
of the adsorption isotherm: the vibrational frequencies of the 
molecule, the adsorption energy, the energy of interaction 
between nearest neighbors and that between next-nearest 
neighbors.
 
\section{The calculation of the adsorption isotherm}
We are now ready to calculate the 
coverage  of the adsorbed layer in equilibrium
with a gas of CO molecules. This calculation is performed
in two ways. The first method uses the equilibrium condition: 
 the chemical potential
of a CO molecule in gas is equal to the chemical potential of an
adsorbed CO molecule. To calculate the latter we use the quasichemical
 approximation\cite{hill}.
The second method, uses Grand Canonical Monte Carlo simulations with the 
chemical potential of the gas calculated from the
equations provided by Statistical Mechanics for ideal gases. 

\subsection{The quasichemical approximation}
The quasichemical approximation gives for 
the chemical potential of CO on Pd
\begin{eqnarray}
\mu_a&=&\varepsilon_a+2\varepsilon_{nn}+\sum_{i=1}^6[\frac{\omega_i}{2}
+k_BT\ln(1-e^{\omega_i/k_BT})]\nonumber\\
\label{eq:muad}
&&\mbox{}+k_BT\ln[(\frac{1-\rho}{\rho})^3(\frac{\rho-a}{1-\rho-a})^2]\ ,
\end{eqnarray}
where   $\rho$ is the coverage of CO (number of molecules divided by the 
number of lattice sites).\cite{hill}  The symbol $a$ stands for
\[a=2\rho(1-\rho)/(\beta+1)\] with  \[\beta=\sqrt{1-4\rho(1-\rho)(1-e^{
-\varepsilon_{nn}/k_BT})}\ .\]

Since the pressure is low, we can use the ideal gas expression for the 
chemical potential\cite{hill} of the gas:
\begin{equation}
\label{eq:mug}
\mu_g=k_BT\ln[Pf(T)]
+\frac{\omega_0}{2}+k_BT\ln(1-e^{-\omega_0/k_BT})\ ,
\end{equation}
with
\begin{equation}
f(T)=\frac{1}{k_BT}(\frac{2\pi\hbar^2}{mk_BT})^{3/2}\frac{\hbar^2}{2Ik_BT}\ .
\end{equation}
In these equations, $P$ is the pressure, $m$ is the mass of CO-molecule,
$I$ is the moment of inertia of CO and $\omega_0$ is the
vibration frequency of free CO, 2114\ cm$^{-1}$.

In equilibrium calculations we must ensure that the two chemical potentials
use the same reference energy. For us, this is the electronic
energy of the slab plus 
the electronic  energy of a CO molecule in gas phase. The
zero point vibrational energy of the molecule in the gas
and on the surface is included in the chemical potential.

By making the chemical potential for the adsorbed CO, [Eq.\ (\ref{eq:muad})],
equal to the chemical potential of the gas [Eq.\ (\ref{eq:mug})],
we obtain a relationship between  
surface coverage $\rho$ and gas
pressure $P$ (the adsorption isotherm). Since we know the
dependence of the parameters on strain, we can calculate
the adsorption isotherm for a strain of $\pm2$\%, at $300$~K.
\begin{figure}[htb]
\psfig{figure=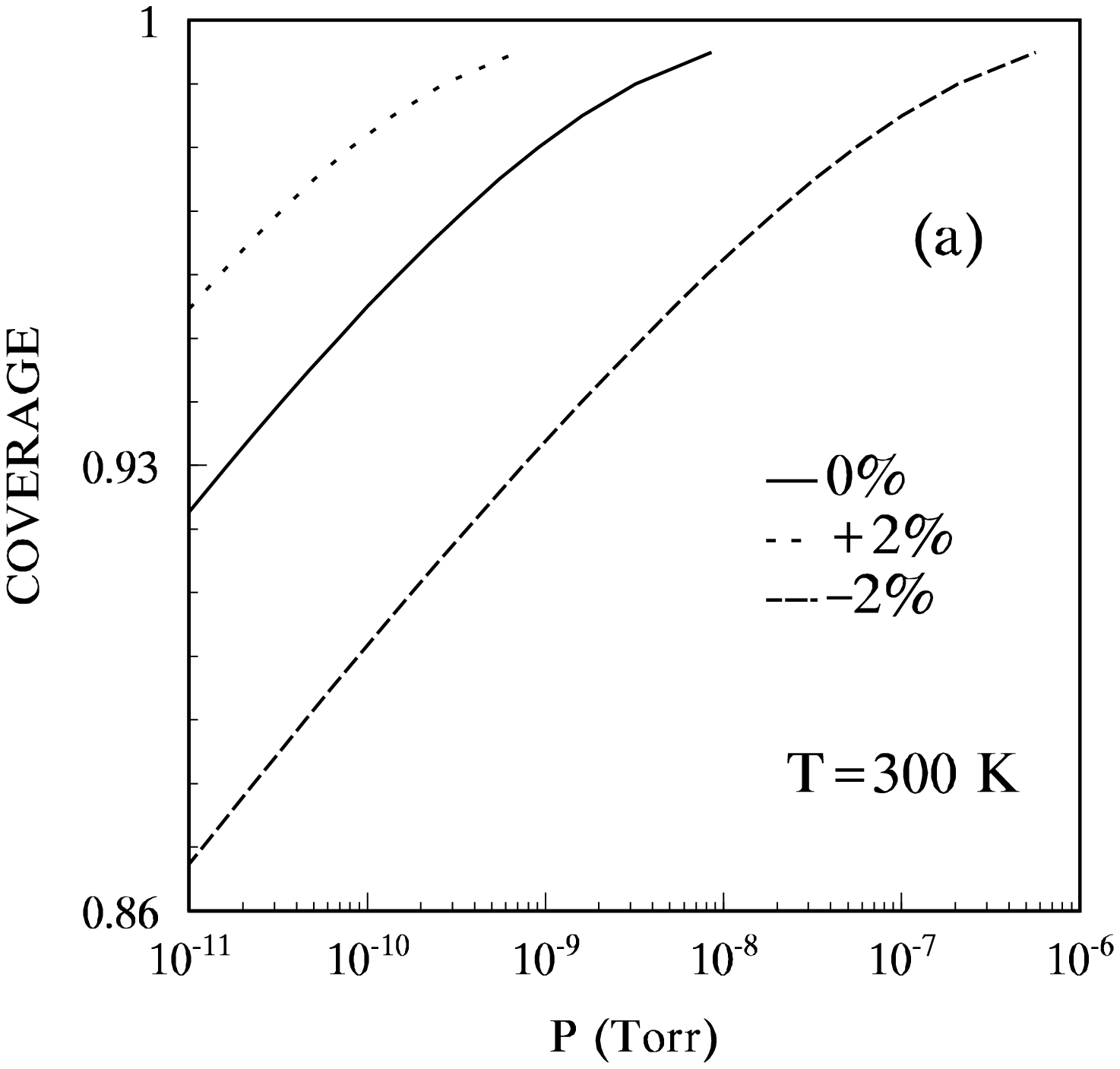,width=8.5cm,height=7.5cm,angle=0}
\psfig{figure=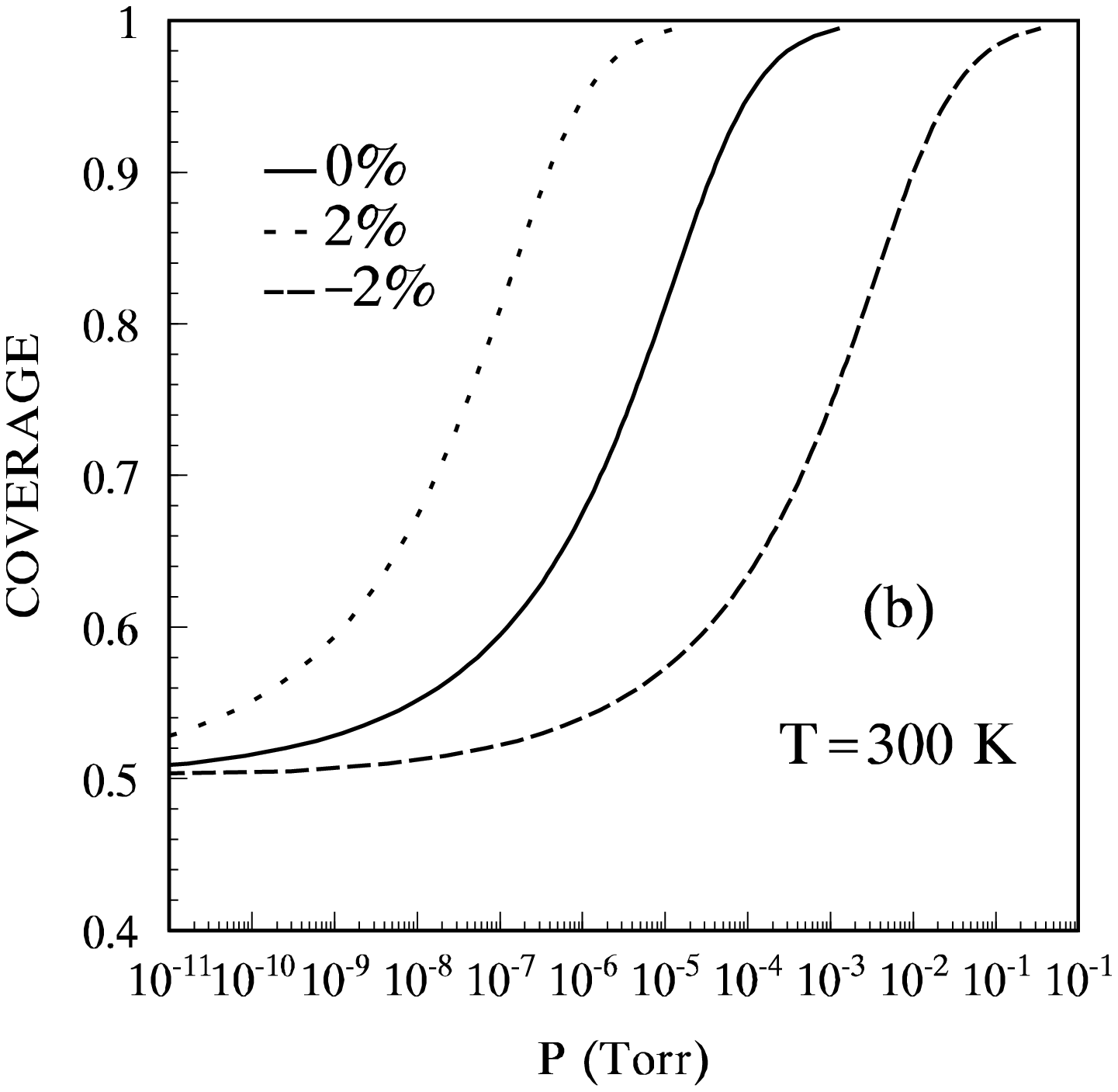,width=8.5cm,height=7.5cm,angle=0}
\caption{Coverage as a function of pressure $P$ for different strains.
( Solid curve: 0
strain; dotted curve: 2\% strain; dashed
curve: $-2$\% strain. $T=300$\ K.
Fig.\ 3(a): Model I, Fig.\ 3(b): Model II.)}
\end{figure}

We first consider Model I, which assumes  that
$\varepsilon_{nn}=\varepsilon_{nnn}=0$ and includes the 
effect of the interactions between the molecules in the
adsorption energy $\bar{\varepsilon}_a$. The magnitude
of $\bar{\varepsilon}_a$ is given in Table\ I, for different coverages
and strains. The values at other coverages are obtained by
interpolation. For the vibrational frequencies 
we use the mean of the values given in Table II\@. 

The adsorption isotherm for Model I is calculated by setting $\mu_g$ given by
 Eq.\ (\ref{eq:mug}) equal to $\mu_a$ given by Eq.\ (\ref{eq:muad}),
with $\epsilon_{nn}=0$ and $\epsilon_a$ replaced with $\bar{\epsilon}_a$.
This gives us an equation connecting the gas pressure to the coverage
(the adsorption isotherm). The dependence of coverage on the gas pressure
obtained with this method  is plotted in Fig.\ 3(a),
for 0  and $\pm 2$~\% strains, at $T=300$\ K. 

In the next calculation we use Model II, which assumes that  the
adsorption energy $\epsilon_a$ is independent of coverage, that the
nearest-neighbors interact with the energy $\epsilon_{nn}$, and
that $\epsilon_{nnn}=0$. The isotherm is obtained by making $\mu_a =\mu_g$.
The resulting adsorption isotherm is plotted in  Fig.\ 3(b).

It is clear that in both models the strain has a very large effect on chemisorption.
For example, the Fig.\ 3(b) shows that  at $10^{-6}$\ Torr the CO coverage on the unstrained
surface is $\sim 0.64$\ ML. Stretching the metal to increase the
lattice constant by $2$\% changes the coverage to $\sim 0.92$\ ML. Another 
way to look at this is to note that a coverage of 0.7 is
achieved at roughly $\sim 1.2\times 10^{-8}$\ Torr if the lattice constant is
increased by $2$\%, at $\sim 1.2 \times 10^{-6}$\ Torr if the surface is not 
strained, and at $\sim 10^{-3}$\ Torr if the lattice constant is
decreased by $2$\%. These effects are large enough to be
easily detected experimentally, if one could find a convenient way
to strain the surface in ultra-high vacuum.

\subsection{Monte Carlo simulation}

We use grand-canonical Monte Carlo (MC)
calculations to determine the coverage of CO on the Pd(100) surface,
 at given gas pressure and temperature. The methodology is 
presented well in an excellent book by Frenkel and Smit\cite{frenkel}
so there is no need to discuss it here.
 The Hamiltonian consists of 
the adsorption energy $\varepsilon_a$ multiplied by the number of 
adsorbed molecules, plus the pairwise interaction energy $\varepsilon_{nn}$
multiplied by the number distinct 
nearest-neigbor  pairs, plus the energy $\varepsilon_{nnn}$ multiplied
by the number of distinct next-nearest-neighbor pairs, plus the adsorption 
energy, plus the vibrational energies of the adsorbate. 
The chemical potential of the gas is calculated from Eq.\ (\ref{eq:mug}).
We use  $1000\times 1000$ lattice sites.

The results of the simulations are plotted  in Fig.\ 4 for  $T=300$\ K.
As in the calculations performed in the previous section, we find that
strain  substantially affects the surface coverage. For example, at a
pressure of $10^{-4}$ Torr and a temperature of 300 K, the coverage is
$\sim~ 0.70$ for $-2~\%$ strain, $\sim~ 0.66$ in the absence of strain, 
and $\sim 0.57$ for $2~\%$ strain. The effect is much stronger if the quasi-chemical
approximation is used (see Fig.\  3 (b)), but we attribute this to
inaccuracies in Model II.  The discrepancy between this model and the
Monte Carlo simulation is so great that we have to discount Model II as
insufficiently accurate. We suspect that this is due to the neglect of
the repulsive,  next-nearest-neighbor interactions, which makes the coverage given by Model II larger than in reality (here the Monte Carlo simulation is the ``reality'').
Model I is so far off from both the quasichemical approximation and the
Monte Carlo simulation that it should be completely discarded. This is a
pity  since, had it been correct, this model  would have allowed a
very simple interpretation of the experiments. 

\begin{figure}[htb]
\psfig{figure=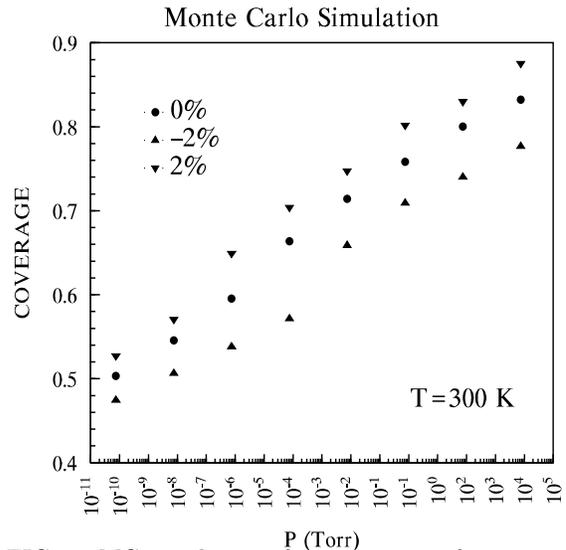,width=8.5cm,height=7.5cm,angle=0}
\caption{MC simulation of coverage as a function of pressure $P$ for
different strains under the assumption of no nn interaction between COs.
$T=300$\ K.}
\end{figure}

\section{Summary}
We have studied how uniform strain in the plane parallel to the (100)
surface of palladium affects the properties of chemisorbed CO. The effect
on the vibrational frequencies is small: it is comparable to or less than
that observed for the molecule located in different binding sites. The
shift in the stretch frequency of CO is sufficiently large to be detected  by
 infrared
spectroscopy.  The effect on the binding energy at low coverage is larger.
We found the weakest  binding when the surface is compressed ({\em  i.e.}
for $-2$~\% strain). Releasing the compressive strain raises the binding energy by
0.025 eV. Going to a tensile strain of $2\%$ increases the binding energy
by another 0.027 eV. From the equilibrium condition (
$\mu_g$ given by  Eq.\@(\ref{eq:mug}) is equal to $\mu_{ad}$ given by
 Eq.\@(\ref{eq:muad}) ) we see that the logarithm of the equilibrium
 pressure is proportional to the adsorption energy. This means that small
 changes in the adsorption energy lead to large changes in
the equilibrium pressure. While the quasi-chemical approximation is not
quantitatively accurate, this particular prediction is reliable. It
is indeed confirmed by the Monte Carlo simulations.

It is interesting to speculate on the broader consequences
of these findings. Ibach's experiments\cite{ibach1,ibach2,ibach3,ibach4}
showed that chemisorption induces surface strain, which results in the
bending of the thin slab on which the adsorption was performed. This
means that in all equilibrium studies we ought to take strain as
an additional  thermodynamic variable\cite{gibbs} and stress as its conjugate variable.
It follows that all equilibria will depend on the strain state of the
surface. Moreover, the outcome of  transformations performed at constant strain will be different
from  that of transformations taking place at constant stress, just as
transformations at constant volume differ from those at constant pressure.
This will affect the 
adsorption isotherm and  chemical equilibrium at the surface.
Since strain affects equilibrium, detailed balance tells us that it must  also affect kinetics. The
only question is whether the modifications caused by strain  are large
enough to make it necessary to include them in a thermodynamic analysis of surface phenomena. The present
work suggests that they are,  and therefore  they deserve further study.

Besides using simple mechanical means, one could induce strain by
passing sound through the solid. In particular, the excitation of the
Rayleigh mode would seem most effective, since it affects the surface
more than other modes. Based on our calculations one would expect
desorption to occur from the regions compressed by the sound wave.
Moreover, chemical equilibrium on the surface is likely to be affected. 
Indeed,  Krishner and Lichtman\cite{licht} 
have shown that passing sound through a surface causes desorption of the molecules 
adsorbed on it. Inoue {\em et al.} \cite{ino1,ino2,ino3,ino4} have demonstrated that  
sound affects the yield of catalytic reactions. Prompted by these experiments King {\em et al.}\cite{king1,king2} have performed careful ultra-high vacuum experiments in which
 the interplay between sound and chemisorbed molecules can be studied in a 
clean and controlled environment. They found that exposing a Pt surface to sound 
can change the rate of CO oxidation. By using photo-electron microscopy, Kelling,
 Cerasari, Rotermundt, Ertl and King \cite{ertl} have shown that the reaction is 
affected mostly by controlling CO adsorption at given CO  pressure. 

In most ``sonochemistry'',  sound acts by heating and compressing the system\cite{sus}. 
This is not the case in the surface science experiments mentioned above, which have 
ruled out heating effects\cite{king1,king2,ertl}. The effect of sound on
 chemical reaction is very hard to explain by conventional arguments: sound
 frequency is very low and excites phonons of long wavelength with very 
low energy. It is unlikely that these can affect the rate of chemical processes. For this 
reason we speculate that perhaps it is the strain induced by the sound wave that 
plays a role in the process. The fact that the presence of sound does not affect the 
activation energy of the oxidation reaction but influences the CO coverage supports t
his view. 

We expect that strain effects also play  a role in supported catalysts, especially 
those using small clusters.  When such a cluster is deposited on a support it 
suffers two modifications: a charge rearrangement and a distortion of its geometry (strain). 
It is therefore possible that a part  of the change in  the catalytic activity cause by
 depositing clusters on a inert support comes from the strain.  

\acknowledgements
We thank Nick  Blake, Ross  Larsen and Professor Jurgen Hafner for help with the
 computations. We are grateful  to Jens Norskov for many useful discussions in the 
early stages of this work. This work has been supported by AFOSR F49620-98-1-0366,  by NSF
 (through CDA96-01954), and by Silicon Graphics Inc.

\end{multicols}

\references
\bibitem[*]{byline} Author to whom correspondence should be
addressed. Email: metiu@chem.ucsb.edu. 

\bibitem{webb} W. E. Packard and M. B. Webb, Phys. Rev. Lett. {\bf 61}, 2469 (1988)
\bibitem{lagally} M. B. Webb, F. K. Men, B. S. Swartzentruber, R. Kariotis and M. G. Lagally, in {\it Kinetics of Ordering and Growth at Surfaces}, ed. M. G.Lagally,(Plenum, New York, 1989)

\bibitem{ibach1} D. Sander and H. Ibach, Phys. Rev B {\bf 43}, 4263 (1991)
\bibitem{ibach2} A. Grossmann, W. Erley and H. Ibach, Surface. Sci. {\bf 313}, 209 (1994)
\bibitem{ibach3} H. Ibach, J. Vac. Sci. Technol. A {\bf 12}, 2440 (1994)
\bibitem{ibach4} A. Grossmann, W. Earley, J. B. Hannon and H. Ibach, Phys. Rev. Lett. {\bf 77}, 127 (1996)
\bibitem{menzel} M. Gsell, P. Jakob and D. Menzel, Science {\bf 280}, 451 (1998)
\bibitem{goodman} J. A. Rodriguez and D. W. Goodman, Science {\bf 257}, 897 (1992)
\bibitem{madey} T. E. Madey, C.-H. Nien, K. Pelhos, J. J. Kolodziej, I M. Abdelrehim, H. -S. Tao, Surface Sci. {\bf 438}, 191 (1999)
\bibitem{behm} F. Buatier de Mongeot, M. Scherer, G. Gleich, E. Kopatzki, and J. 
Behm, Surface Sci. {\bf 411}, 249 (1998)
\bibitem{kern} E. Kampshoff, E. Hahn, K. Kern, Phys. Rev. Lett. {\bf 73}, 704 (1994)
\bibitem{chork} J. H. Larsen and I. Chorkendorff, Surface Sci. {\bf 405}, 62 (1998)
\bibitem{besenbacher} M. \O.  Pedersen,  S. Helveg, A. Ruban, A. Stensgaard, E. L\ae gsgaard, J. K. Norscov, and F. Bessenbacher, Surface Sci. {\bf 426}, 395 (1999)
\bibitem{scheffler} C. Ratsch, A. P. Seitsonen and M. Scheffler, Phys. Rev. B{\bf 55}, 6750 (1997) 
\bibitem{metiu1} T. R. Mattsson and H. Metiu, Appl. Phys. Lett. {\bf 75}, 926 (1999)
\bibitem{petroff} A. E. Romanov, P. M. Petroff and J. S. Speck, Appl. Phys. Lett. {\bf 74}, 2280 (1999)
\bibitem{norskov} M. Mavrikakis, B. Hammer and J. K. N\o rskov, Phys. Rev. Lett. {\bf 81}, 2819 (1998)
\bibitem{ibachreview} H. Ibach, Surface Sci. Reports  {\bf 29}, 193 (1997)
\bibitem{norskovreview} J. Norskov, Advances in Catalysis, {\bf 43}, xxxx (2000)
\bibitem{kress1} G. Kresse and J. Hafner, Phys. Rev. B {\bf 47}, 558 (1993);
{\bf 49}, 14251 (1994)
\bibitem{kress2} G. Kresse and J. Furthm\"uller, Phys. Rev. B {\bf 54}, 11169
(1996); Comput. Mater. Sci. {\bf 6}, 15 (1996)
\bibitem{van} D. Vanderbilt, Phys. Rev. B {\bf 41}, 7892 (1990); G. Kresse
and J. Hafner, J. Phys.: Condens. Matter {\bf 6}, 8245 (1996)
\bibitem{wang} Y. Wang and J.P. Perdew, Phys. Rev. B {\bf 44}, 13928 (1991)
\bibitem{eichler} A. Eichler and J. Hafner, Phys. Rev. B {\bf 57}, 10110
(1998)
\bibitem{frenkel} D. Frenkel and B. Smit, {\it Understanding Molecular
Simulation} (Academic Press, San Diego, 1996)
\bibitem{hill} T. Hill,  {\it Statistical Mechanics} (Dover Publicatons, New
York, 1956), p. 348
\bibitem{gibbs} {\it The Scientific Papers of J. Willard Gibbs }, vol. 1,
Thermodynamics, (Dover Publications, New York, 1961), p. 184
\bibitem{licht} C. Krishner and D. Lichtman, Phys. Lett. {\bf 44}A, 99 (1973)
\bibitem{ino1} Y. Inoue, Y. Matsukawa and  K. Sato, J. Am. Chem. Soc. {\bf 111}, 8965 (1989)
\bibitem{ino2} Y. Inoue, Y. Matsukawa and K. Sato, J. Phys. Chem. {\bf 96}, 2222 (1992)
\bibitem{ino3} Y. Inoue and Y. Watanabe, Catal. Today {\bf 16}, 487 (1993)
\bibitem{ino4} Y. Inoue, J. Chem. Soc., Faraday Trans. {\bf 90}, 815 (1994)
\bibitem{ino5} Y. Watanabe, Y. Inoue and K. Sato, Surf. Sci. {\bf 358}, 769 (1996)
\bibitem{king1} M. Gruyters, T. Mitrelias and D.A. King, Appl. Phys. A {\bf 61}, 243 (1995)
\bibitem{king2} S. Kelling, T. Mitrelias, Y. Matsumoto, V. P. Ostanin and D. A. King, J. 
Chem. Phys. {\bf 107}, 5609 (1997)
\bibitem{ertl} S. Kelling, S. Cesari, H. H. Rotermund, G. Ertl and D. A. King, Chem. Phys. Lett. {\bf 293}, 325 (1998)
\bibitem{sus} K. S. Suslik, Science {\bf 247}, 1439 (1990)
\end{document}